\documentclass[10pt]{article}

\usepackage{amsmath}
\usepackage{amssymb}

\usepackage{graphicx}

\usepackage{cite}

\usepackage{color} 
\usepackage{mathtools}
\usepackage{amsthm}
\usepackage{array}
\usepackage{mathrsfs}

\usepackage[labelfont=bf,labelsep=period,justification=raggedright]{caption}

\makeatletter
\renewcommand{\@biblabel}[1]{\quad#1.}
\makeatother

\date{\today}

\newtheorem{defn}{Definition}
\newtheorem{theorem}{Theorem}
\newtheorem{lemma}[theorem]{Lemma}
\newtheorem{cor}[theorem]{Corollary}
\newtheorem{remark}[theorem]{Remark}

\begin{document}

\begin{flushleft}
{\Large
\textbf{Information Inequalities and Generalized Graph Entropies}
}
\\
Lavanya Sivakumar, Matthias Dehmer$^{1,\ast}$
\\
{\bf 1} Institute for Bioinformatics and Translational Research,
UMIT, Hall in Tyrol, Austria
\\
$\ast$ E-mail: matthias.dehmer@umit.at
\end{flushleft}

\section*{Abstract}
In this article, we discuss the problem of establishing relations between information measures assessed for network structures. Two types of entropy based measures namely, the Shannon entropy and its generalization, the R\'{e}nyi entropy have been considered for this study. Our main results involve establishing formal relationship, in the form of implicit inequalities, between these two kinds of measures when defined for graphs. Further, we also state and prove inequalities connecting the classical partition-based graph entropies and the functional-based entropy measures. In addition, several explicit inequalities are derived for special classes of graphs. 

\noindent \textbf{Keywords: }
Information Measures, Information inequalities, Topological complexity, Shannon Entropy, R\'{e}nyi Entropy, Networks. 

\section*{Introduction}
\label{sec:Intro}


Complexity of a system, in general, deals with the intricate design and complex interrelations among the components of the system. One of the plausible categorization of  complexity analysis is based on the functional behavior, topological properties, and/or at the compositional level of a system \cite{chLS:Bonchev-Rouvray1}. Over the years, all these categories have been implemented and contemplated concurrently in all branches of science and social science. 
However, here we restrict our attention to the topological complexity of network-based systems.This relates to determining the complexity of the underlying graph structure of a network. 
The quantitative estimation (using measures/indices) of topological complexity has been proven useful in characterizing the networks and has widely spread into all branches of natural sciences, mathematics, statistics, economics and sociology; for e.g., see \cite{chLS:Anand-Bianconi, chLS:Costa1, chLS:Kim-Wilhelm, ChLS:Balaban2, chLS:Bertz, chLS:Basak, chLS:Bonchev-Rouvray, chLS:Claussen, chLS:Korner, ChLS:Butts}. 
In the study of complexity, information theory has been playing a predominant role. That is, the measures based on Shannon entropy has been very powerful and useful in determining the topological complexity of networks; see \cite{bonchev_2009,bonchev_2,chLS:Bonchev-Rouvray1,chLS:Claussen}.  Apart from Shannon entropy, its generalizations such as R\'{e}nyi entropy \cite{renyi_1961}, Dar\`{o}czy entropy \cite{daroczy_1979} have also been identified as potential contributors in characterizing network-based systems; see \cite{chLS:Dehmer3}. 

In this paper, we deal with an interesting and significant aspect when analyzing the complexity of network-based systems. Namely, we establish relations between information-theoretic complexity measures \cite{SL-Dehmer1,chLS:Dehmer-Mowshowitz-Survey}. Investigating relations (in the form of inequalities) among measures, by and large is very useful when studying large scale networks where evaluating the exact value of a measure is computationally hard and time-consuming. In addition, they also serve as a tool for solving problems. In communication theory, inequalities have paved way to the development of so-called algebra of information, various rules involving mutual information and entropy measure. For example, inequalities such as Young's inequality, Brunn-Minkowski inequality, Fisher's information inequalities have been main contributors to the mentioned development \cite{Dembo-Cover-Thomas,yeung_1,dragomir_1997}. 

Inequalities involving information measures are also referred to as {\it information inequalities} \cite{chLS:Dehmer-Mowshowitz1}.
They are  classified in two types, namely the {\it implicit information inequalities}  and {\it explicit information inequalities}. In particular, when information measures are present on either side of the inequality, we call it an \textit{implicit information inequality}\cite{chLS:Dehmer-Mowshowitz1}, while in the latter, the information measure is bounded by a function of parameters (or constants) involved.
For some of the recent works in this direction, we refer to \cite{chLS:Dehmer-Mowshowitz1,chLS:Bonchev-Trinajstic,dehmer_mowshowitz_emmert_1_2011,SL-Dehmer1,chLS:Dehmer-Streib5}.

The main contribution of this paper, is to establish implicit information inequalities involving the Shannon entropy and the R\'{e}nyi entropy measures when applied to network graphs. Secondly, we also present implicit inequalities between R\'{e}nyi entropy measures having two different types of probability distributions with additional assumptions. 
To achieve this, we analyze and establish the relation between classical partition-based graph entropies \cite{bonchev_Trinajstic_1977,bonchev_2,mowshowitz2009} and the non-partition based (or the functional) based entropies \cite{chLS:Dehmer2}. Finally, we apply the obtained inequalities to specific graph classes and derive simple explicit bounds for the R\'{e}nyi entropy. 

\section*{Methods}

In this section, we state some of the definitions of information-theoretic complexity measures \cite{chLS:Dehmer1,dehmer_mowshowitz_2010,skorobogatov,chLS:Shannon}. Measures are based on two major classifications, namely partition based and partition-independent measures. Some basic results on inequalities on real numbers \cite{hardy_1988,dragomir_1997} are also presented towards the end of the section. 

Let $G=(V,E)$ be a graph on $N$ vertices where $V = \{v_1, v_2, \ldots, v_N \}$ and $E \subseteq V \times V$. Throughout this article, $G$ denotes a simple undirected graph. Let $X$ be a collection of subsets of $G$ representing a graph object. Let $\Gamma$ be an equivalence relation that partitions $X$ into $k$ subsets $X_1, X_2, \ldots, X_k$, with cardinality $|X_i|$, for $1 \leq i \leq k$.  Let $\{p_1, p_2, \ldots, p_k\}$ denote the probability distribution on $X$ w.r.t $\Gamma$, such that $p_i = \displaystyle \frac{|X_i|}{|X|}$ $(1\leq i \leq k)$, is the value of probability on each of the partition. 

For graphs, the Shannon's entropy measure \cite{chLS:Shannon1} is also referred to as ``the information content of graphs" \cite{chLS:Rashevsky, chLS:Bonchev} and is defined as follows:  
\begin{defn}
The mean information content,  $H_\Gamma(G)$,  of $G$ with respect to $\Gamma$ is given by 
\begin{eqnarray}
H_\Gamma(G) = &- \displaystyle \sum_{i=1}^k p_i \log_2 p_i  =& - \displaystyle \sum_{i=1}^k \frac{|X_i|}{|X|} \log_2 \frac{|X_i|}{|X|}.
\label{Shannon-Entropy-2}
\end{eqnarray}
\end{defn}
Note that while the above definition is based on partitioning a graph object, another class of Shannon entropy has been defined by \cite{chLS:Dehmer1} where the probability distribution is independent of partitions. That is, probabilities were defined for every vertex of the graph using the concept of information functionals. 

Suppose $f: V \rightarrow R^{+}$ is an arbitrary information functional \cite{chLS:Dehmer1} that maps a set of vertices to the non-negative real numbers and let  
\begin{eqnarray}\label{Prob-funtional}
p(v) & = & \displaystyle\frac{f(v)}{\sum_{v \in V}f(v)},
\end{eqnarray}
denote the probability value of $v \in V$. Here, 

\begin{defn}
The graph entropy, $H_f(G)$, representing the structural information content of $G$ {\rm \cite{chLS:Dehmer1,dehmer_mowshowitz_2010}} is then given by, 
\begin{equation}\label{EntropyEquation-defnFnal}
H_f(G) = -\displaystyle \sum_{i = 1}^N p(v_i) \log_2 p(v_i) = -\displaystyle \sum_{i = 1}^N  \frac{f(v_i)}{\sum_{j=1}^N f(v_j)} \log_2 \frac{f(v_i)}{\sum_{j=1}^N f(v_j)}.
\end{equation}
\end{defn}

As a follow-up to Shannon's  seminal work \cite{chLS:Shannon}, many generalizations of the entropy measure were proposed in the literature \cite{renyi_1961,daroczy_1979,arndt_2004}. These generalized entropies were recently  \cite{chLS:Dehmer3}, extended to study the class of graphs. In the following, we present one such generalization from \cite{chLS:Dehmer3}, namely the R\'{e}nyi entropy for graphs.  
\begin{defn}
The R\'{e}nyi entropy $H_{\alpha,\Gamma}(G)$, for $0 < \alpha <\infty$ and $\alpha \neq 1$,  of a graph $G$ {\rm \cite{chLS:Dehmer3}} is given by,
\begin{equation}\label{RenyiEntropyEquation-defn}
H_{\alpha,\Gamma}(G) = \displaystyle \frac{1}{1-\alpha} \log_2\left(\sum_{i = 1}^k (p_i)^\alpha\right) = \displaystyle \frac{1}{1-\alpha} \log_2\left(\sum_{i=1}^k \left(\frac{|X_i|}{|X|}\right)^\alpha \right).
\end{equation}
Here, $\Gamma$ is the equivalence relation on a graph object and $p_i$ $(1 \leq i \leq k)$ denotes the probabilities defined on the partition induced by $\Gamma$.  
\end{defn}

It has been proved that R\'{e}nyi entropy is a generalization of Shannon entropy and in the limiting case when $\alpha \to 1$, the R\'{e}nyi entropy equals the Shannon entropy \cite{arndt_2004}. 

Similar to expression \eqref{EntropyEquation-defnFnal}, the R\'{e}nyi entropy can be immediately extended\cite{chLS:Dehmer3} to partition-independent probability distributions defined on $G$. 
\begin{defn}
Let $H_{\alpha,f}(G)$, for $0 < \alpha <\infty$ and $\alpha \neq 1$, denote the R\'{e}nyi entropy {\rm \cite{chLS:Dehmer3}}defined using an information functional $f$. Then 
\begin{equation}
H_{\alpha,f}(G) = \displaystyle \frac{1}{1-\alpha} \log_2\left(\sum_{i = 1}^N (p(v_i))^\alpha\right) = \displaystyle \frac{1}{1-\alpha} \log_2\left(\sum_{i = 1}^N \left[\frac{f(v_i)}{\sum_{j=1}^N f(v_j)}\right]^\alpha\right).
\label{RenyiEntropyEquation-defnFnal}
\end{equation}
\end{defn}
Next we state some interesting inequalities from the literature that are crucial to prove our main results.
One of the well-known result for the real numbers is stated as follows \cite{hardy_1988}.
\begin{lemma}\label{Lemma1}{\rm\cite{hardy_1988}}
Let $x,y > 0$ and $x \neq y$ be real numbers. Then 
\begin{eqnarray}
ry^{r-1}(x-y) &< x^r - y^r <& rx^{r-1}(x-y), \label{Fund-ineq1}
\shortintertext{ if  $r<0$  or $r>1$,} 
rx^{r-1}(x-y) &< x^r - y^r <& ry^{r-1}(x-y), 
\label{Fund-ineq2}
\end{eqnarray}
if $0 < r < 1.$ ~\hfill{$\Box$}
\end{lemma}
A simplified form of Minkowski's inequality is given by \cite{hardy_1988}: 
\begin{lemma}\label{Lemma2}{\rm\cite{hardy_1988}}
If $r>0$, then 
\begin{equation}
(\sum_{i}(a_i +b_i +\cdots +l_i)^r)^R \leq (\sum_{i}(a_i)^r)^R +(\sum_{i}(b_i)^r)^R +\cdots +(\sum_{i}(l_i)^r)^R
\end{equation}
where $R= 1$, if $0 < r \leq 1$ and $R= \frac{1}{r}$, if $r > 1$. \hfill{$\Box$}
\end{lemma}
As an extension of discrete Jensen's inequality, the following inequality was derived in \cite{dragomir_1997}. 
\begin{lemma}\label{Lemma3} {\rm\cite{dragomir_1997}}
Let $x_k \in (0,\infty)$, for $1\leq k \leq n$, and  $p_k \geq 0$ such that $\sum_{k =1}^{n} p_k=1$. Then
\begin{equation}
0 \leq \log_2 \left(\sum_{k=1}^n p_k x_k\right) - \sum_{k=1}^n p_k \log_2 x_k \leq \frac{1}{2 \ln 2} \sum_{k,i =1}^n \frac{p_k p_i}{x_k x_i}(x_i - x_k)^2.
\label{Ineq-Dragomir}
\end{equation}
~\hfill{$\Box$}
\end{lemma}

\section*{Results}
In this section, we present our main results on implicit information inequalities. To begin with, we establish the bounds for  R\'{e}nyi entropy in terms of Shannon entropy. 

\begin{theorem}\label{Thm-alpha-f-f}
Let $p(v_1), p(v_2), \ldots, p(v_N)$ be the probability values on the vertices of a graph $G$. Then the R\'{e}nyi entropy can be bounded by the Shannon entropy as follows:\\
When $0 <\alpha < 1$,
\begin{equation}
H_f(G) \leq H_{\alpha,f}(G)< H_f(G) + \frac{N(N-1)(1-\alpha)\rho^{\alpha-2}}{2\ln 2}. 
\label{Eqn1}
\end{equation}
When $\alpha >1$, 
\begin{equation}
H_f(G) -  \frac{(\alpha-1)N(N-1)}{2\ln 2\cdot \rho^{\alpha-2}} < H_{\alpha,f}(G)\leq H_f(G), 
\label{Eqn2}
\end{equation}
where $\rho = \displaystyle \max_{i,k}\frac{p(v_i)}{p(v_k)}$. 
\end{theorem}

\noindent {\bf Proof:} It is well known \cite{arndt_2004} that the R\'{e}nyi entropy satisfies the following relation with the Shannon entropy .
\begin{eqnarray}
H_{\alpha,f}(G) &\geq & H_f(G), \text{ if } 0 < \alpha <1,
\shortintertext{and }
H_{\alpha,f}(G) &\leq & H_f(G) , \text{ if } \alpha > 1.
\end{eqnarray}

\noindent Next let us prove the other bound of $H_{\alpha,f}(G)$. \\
Let $\rho = \displaystyle \max_{i,k}\frac{p(v_i)}{p(v_k)}$.
Consider, the inequality \eqref{Ineq-Dragomir} from Lemma \ref{Lemma3} with $p_k = p(v_k)$ and $x_k = p(v_k)^{\alpha-1}$.
We get, 
\begin{equation}
\log_2 \left(\sum_{k=1}^N p(v_k)^\alpha\right) - (\alpha - 1)\sum_{k=1}^N p(v_k) \log_2 p(v_k) 
\leq   \frac{1}{2\ln 2} \sum_{i, k=1}^N \frac{p(v_k)p(v_i)}{(p(v_k)p(v_i))^{\alpha-1}}(p(v_i)^{\alpha-1}-p(v_k)^{\alpha-1})^2. 
\label{Eqn-inter1}
\end{equation}
Now we prove the theorem by considering intervals for $\alpha$. 

\noindent {\bf Case 1:} When $0 <\alpha <1$.
 
Dividing by $(1-\alpha)$ on either side of the expression \eqref{Eqn-inter1}, we get 
\begin{equation}
H_{\alpha,f}(G) -  H_f(G) \leq  \frac{1}{2\ln 2 (1- \alpha)} \sum_{i, k=1}^N \frac{p(v_k)p(v_i)}{(p(v_k)p(v_k))^{\alpha-1}}(p(v_i)^{\alpha-1}-p(v_k)^{\alpha-1})^2. 
\label{Eqn-inter2}
\end{equation}
Using inequality \eqref{Fund-ineq1} from Lemma \ref{Lemma1}, with $r = \alpha -1 < 0$ in the following sum we get, 
\begin{eqnarray}
\displaystyle\sum_{k,i=1}^N \frac{(p(v_i)^{\alpha-1}-p(v_k)^{\alpha-1})^2 }{(p(v_i)p(v_k))^{\alpha-2}}&<& \sum_{\begin{subarray}{c}k,i=1 \\ i \neq k\end{subarray}}^N \frac{(\alpha - 1)^2 p(v_i)^{\alpha-2} (p(v_i)-p(v_k))^2}{p(v_k)^{\alpha-2}}, \\ 
&\leq& \sum_{\begin{subarray}{c}k,i=1 \\ i \neq k\end{subarray}}^N [(\alpha -1)(p(v_i)-p(v_k))]^2 \rho^{\alpha -2} ~(\text{since }\rho:= \max_{i,k}\frac{p(v_i)}{p(v_k)}),\\ 
&<& \sum^N_{\begin{subarray}{c}k,i=1 \\ i \neq k\end{subarray}} (\alpha -1)^2\rho^{\alpha -2}~(\text{since } p(v_i)-p(v_k) <1), \\
&=& \rho^{\alpha -2}(\alpha -1)^2 N(N-1).
\end{eqnarray}
Thus, expression \eqref{Eqn-inter2} becomes
\begin{equation*}
H_{\alpha,f}(G) -  H_f(G) < \frac{\rho^{\alpha -2}(1-\alpha)N(N-1)}{2\ln 2}.
\end{equation*}
which is the required upper bound in \eqref{Eqn1}.

\noindent {\bf Case 2:} When $ \alpha >1$. 

In this case dividing by $(1-\alpha)$ on either side of the expression \eqref{Eqn-inter2}, we get,
\begin{equation}
H_{\alpha,f}(G) -  H_f(G) \geq  \frac{1}{2\ln 2 (1- \alpha)} \sum_{i, k=1}^N \frac{p(v_k)p(v_i)}{(p(v_k)p(v_k))^{\alpha-1}}(p(v_i)^{\alpha-1}-p(v_k)^{\alpha-1})^2. 
\label{Eqn-inter3}
\end{equation}
Using inequality \eqref{Fund-ineq2} with $r = \alpha -1 < 1$ for $1 < \alpha < 2$ and inequality \eqref{Fund-ineq1} with $r = \alpha -1 > 1$, for $\alpha > 2$ we get,
\begin{eqnarray}
\displaystyle\sum_{k,i=1}^N \frac{(p(v_i)^{\alpha-1}-p(v_k)^{\alpha-1})^2 }{(p(v_i)p(v_k))^{\alpha-2}}
&>& \sum_{\begin{subarray}{c}k,i=1 \\ i \neq k\end{subarray}}^N \frac{(\alpha - 1)^2 p(v_i)^{\alpha-2} (p(v_i)-p(v_k))^2}{p(v_k)^{\alpha-2}}, \\ 
& \geq & \sum_{\begin{subarray}{c}k,i=1 \\ i \neq k\end{subarray}}^N \frac{[(\alpha -1)(p(v_i)-p(v_k))]^2}{\rho^{\alpha -2}}, \\ 
&>& \sum^N_{\begin{subarray}{c}k,i=1 \\ i \neq k\end{subarray}} \frac{(\alpha -1)^2}{\rho^{\alpha -2}},~(\text{since } p(v_i)-p(v_k) > -1), \\
&=& \frac{(\alpha -1)^2 N(N-1)}{\rho^{\alpha -2}}.
\end{eqnarray}
Thus, expression \eqref{Eqn-inter3} becomes
\begin{equation*}
H_{\alpha,f}(G)- H_f(G) > \frac{(1-\alpha)N(N-1)}{2\ln 2 \cdot \rho^{\alpha-2}}, 
\end{equation*}
which is the desired upper bound in \eqref{Eqn2}.
\hfill{$\Box$}
%
%

\begin{cor}\label{Cor1-alpha-f-f-epsilon}
In addition, suppose $\epsilon = \displaystyle \max_{i,k} (p(v_i) -p(v_k))$, then 
\begin{eqnarray}
H_f(G) &\leq H_{\alpha,f}(G) < & H_f(G) + \frac{n(n-1)(1-\alpha)\epsilon^2\rho^{\alpha-2}}{2\ln 2}, 
\label{Eqn3}
\shortintertext{when $0<\alpha <1$ and } 
 H_f(G) &\geq H_{\alpha,f}(G) > & H_f(G) -  \frac{(\alpha-1)n(n-1)}{2\ln 2 \cdot \rho^{\alpha-2}}, 
\label{Eqn4}
\end{eqnarray} 
when $\alpha >1$. \hfill{$\Box$}
\end{cor}

\begin{remark}
{\rm Observe that the Theorem \ref{Thm-alpha-f-f}, in general holds for any arbitrary probability distribution with non-zero probability values. The following theorem illustrates this fact with the help of a probability distribution obtained by partitioning a graph object. 
}\end{remark}

\begin{theorem} \label{Thm-alpha-Gamma-Gamma}
Let $p_1, \ldots, p_k$ be the probabilities of the partitions obtained using an equivalence relation $\Gamma$ as stated before. Then 
\begin{eqnarray}
H_\Gamma(G) & \leq  H_{\alpha,\Gamma}(G) < & H_\Gamma(G) + \frac{k(k-1)(1-\alpha)\rho^{\alpha-2}}{2\ln 2}, 
\shortintertext{when $0<\alpha <1$, and }
H_\Gamma(G) & \geq H_{\alpha,\Gamma}(G) > &  H_\Gamma(G) -  \frac{(\alpha-1)k(k-1)}{2\ln 2 \cdot \rho^{\alpha-2}}, 
\end{eqnarray}
when $\alpha > 1$.
\end{theorem} 
\noindent \textbf{Proof:} Proceeding similar to Theorem \ref{Thm-alpha-f-f}, we get the desired result. \hfill{$\Box$}

In the next theorem, we establish bounds between like-entropy measures, by considering the two different probability distributions.

\begin{theorem}\label{Thm-Gamma-f}
Suppose $|X_i| < f(v_i)$, for $1\leq i \leq k$, then 
\begin{eqnarray}
H_{\alpha,\Gamma}(G) &< & H_{\alpha,f}(G) + \frac{\alpha}{1-\alpha} \log \left(\frac{S}{|X|}\right), 
\shortintertext{ if  $0 <\alpha < 1$,} 
H_{\alpha,\Gamma}(G) &> & H_{\alpha,f}(G) - \frac{\alpha}{\alpha-1} \log \left(\frac{S}{|X|}\right),  
\end{eqnarray}
if  $\alpha > 1$. Here $S = \sum_{i =1}^N f(v_i)$. 
\end{theorem}

\noindent {\bf Proof:} Let $S = \sum_{i = 1}^N f(v_i)$ and thus $p(v_i) = \displaystyle \frac{f(v_i)}{S}$ . Now, given $|X_i|< f(v_i)$, for $1 \leq i \leq k$ we have, 
\begin{eqnarray}
\frac{|X_i|}{|X|} &<& \frac{f(v_i)}{|X|} \quad = \quad  \frac{Sp(v_i)}{|X|}.
\label{X_i_rel_f(V)}
\end{eqnarray}
By raising either side of the expression to the power $\alpha$, and by applying summation over $i$ from $1$ to $k$ on either side we get, 
\begin{eqnarray}
\displaystyle \sum_{i =1}^{k}\left(\frac{|X_i|}{|X|}\right)^\alpha &<& \displaystyle \sum_{i =1}^{k}\left(\frac{Sp(v_i)}{|X|}\right)^\alpha, \\
&=& \displaystyle \left(\frac{S}{|X|}\right)^\alpha \sum_{i =1}^{k}(p(v_i))^\alpha.
\shortintertext{Taking logarithms on either side, we obtain  }
\displaystyle \log \sum_{i =1}^{k}\left(\frac{|X_i|}{|X|}\right)^\alpha &<&\displaystyle \log\left(\frac{S}{|X|}\right)^\alpha \sum_{i =1}^{k}(p(v_i))^\alpha, \\
& = & \log\left(\frac{S}{|X|}\right)^\alpha + \displaystyle \log\sum_{i =1}^{k}(p(v_i))^\alpha, \\
& < & \log\left(\frac{S}{|X|}\right)^\alpha + \displaystyle \log\sum_{i =1}^{N}(p(v_i))^\alpha. \label{eqn-inter1}
\end{eqnarray}
Now we make two cases, depending on $\alpha$ as follows: 

\noindent \textbf{Case 1:} When $0<\alpha<1$, dividing by $1-\alpha$ on either side of equation \eqref{eqn-inter1}, we get 
\begin{eqnarray}
H_{\alpha,\Gamma}(G) &< & H_{\alpha,f}(G) + \frac{\alpha}{1-\alpha}\log\frac{S}{|X|}. \label{Result1}
\end{eqnarray}
\textbf{Case 2:} When $\alpha >1$, dividing by $1-\alpha$  on either side of equation \eqref{eqn-inter1}, we get 
\begin{eqnarray}
H_{\alpha,\Gamma}(G) &> & H_{\alpha,f}(G) + \frac{\alpha}{1-\alpha}\log\frac{S}{|X|}. \label{Result2}
\end{eqnarray}
Expressions \eqref{Result1} and \eqref{Result2} are the required inequalities. \hfill{$\Box$}
%
%

\begin{remark}
{\rm A similar relation by considering $H_{\Gamma}(G)$  and $H_{f}(G)$ has been derived in \cite{dehmer_mowshowitz_emmert_1_2011}.
}\end{remark}

We focus our attention to the R\'{e}nyi entropy measure defined using information functionals (given by equation \eqref{RenyiEntropyEquation-defnFnal}) and present various bounds when two different functionals and their probability distributions satisfy certain initial conditions. A similar study has been performed in the case of Shannon entropy; see\cite{SL-Dehmer1,chLS:Dehmer-Mowshowitz1}. 

Let $f_1$ and $f_2$ be two information functionals defined on a graph $G= (V,E)$. Let $S_1 = \sum_{i = 1}^N f_1(v_i)$ and $S_2 = \sum_{i = 1}^N f_2(v_i)$. Let $p_{f_1}(v)$ and $p_{f_2}(v)$ denote the probabilities of $f_1$ and $f_2$, respectively, on a vertex $v \in V$. Let $H_{\alpha,f_1}(G)$ and $H_{\alpha,f_2}(G)$ denote the R\'{e}nyi entropy based on the functionals $f_1$ and $f_2$ respectively.  

\begin{theorem}
Suppose $p_{f_1}(v) \leq \psi * p_{f_2}(v) $, $\forall v \in V$ and $\psi >0$ a constant, then 
\begin{eqnarray}
H_{\alpha,f_1}(G)  & \leq & H_{\alpha,f_2}(G) + \frac{\alpha}{1- \alpha}\log \psi,  \label{Eqn5} 
\shortintertext{ if $0 <\alpha<1$,}
H_{\alpha,f_1}(G)  & \geq & H_{\alpha,f_2}(G) - \frac{\alpha}{\alpha-1}\log \psi,  \label{Eqn6}
\end{eqnarray} 
if $\alpha > 1$.
\end{theorem} 
\noindent  {\bf Proof:} Given  $p_{f_1}(v) \leq \psi * p_{f_2}(v)$. By raising to the power of  $\alpha$ and by applying summation over the vertices of $G$, we get, 
\begin{eqnarray}
\sum_{v \in V} p_{f_1}(v)^\alpha & \leq & \sum_{v \in V} (\psi\cdot p_{f_2}(v))^\alpha.  
\shortintertext{Taking logarithms on either side, we get,  }
\log\left(\sum_{v \in V} p_{f_1}(v)^\alpha\right) &\leq &\log \left(\sum_{v \in V} (\psi\cdot p_{f_2}(v))^\alpha\right), \\ 
& = &\alpha \log \psi + \log \left(\sum_{v \in V} (p_{f_2}(v))^\alpha\right).
\end{eqnarray}
{\bf Case 1:} When $0 <\alpha < 1$. Dividing either side of the equation by $(1- \alpha)$, we get the desired expression \eqref{Eqn5}.

\noindent {\bf Case 2:} When $\alpha > 1$. In this case, dividing either side of the equation by $(1- \alpha)$, we get the expression \eqref{Eqn6} as desired. \hfill{$\Box$}
%
%

\begin{cor}
Suppose $f_1(v) \leq f_2(v)$, $\forall v \in V$, then 
\begin{eqnarray}
H_{\alpha,f_1}(G)  &\leq & H_{\alpha,f_2}(G) + \frac{\alpha}{1- \alpha}\log \frac{S_2}{S_1}, 
\shortintertext{ if  $0 <\alpha<1$, }
H_{\alpha,f_1}(G)  &\geq & H_{\alpha,f_2}(G) - \frac{\alpha}{\alpha-1}\log \frac{S_2}{S_1},
\end{eqnarray}
if  $\alpha > 1$. 
\end{cor}
\noindent{\bf Proof:} By the assumption, we have $p_{f_1}(v) \leq \displaystyle\frac{S_2}{S_1}p_{f_2}(v)$. Therefore, the corollary follows by letting $\psi =  \displaystyle\frac{S_2}{S_1}$ in the above theorem. \hfill{$\Box$}
%

\begin{theorem}
Suppose $p_{f_1}(v) \leq p_{f_2}(v) + \phi$, $\forall v \in V$ and $\phi >0$ a constant, then 
\begin{eqnarray}
H_{\alpha,f_1}(G)  &<& H_{\alpha,f_2}(G) + \frac{1}{1- \alpha}\frac{N\cdot \phi^\alpha}{\sum_{v \in V} (p_{f_2}(v))^\alpha}, \label{Eqn7} 
 \shortintertext{if $0 <\alpha<1$,}
H_{\alpha,f_1}(G)  &>& H_{\alpha,f_2}(G) - \frac{\alpha}{\alpha-1}\cdot \frac{N^{1/\alpha}\cdot \phi}{\left(\sum_{v \in V} (p_{f_2}(v))^\alpha\right)^{1/\alpha}}, \label{Eqn8} 
\end{eqnarray}
if $\alpha > 1$.
\end{theorem}
\noindent {\bf Proof: }Suppose $p_{f_1}(v) \leq p_{f_2}(v) + \phi$, $\forall v \in V$. 
By raising to the power $\alpha$ and by applying summation over the vertices of $G$, we get, 
\begin{equation}
\sum_{v \in V} p_{f_1}(v)^\alpha \leq \sum_{v \in V} (p_{f_2}(v)+ \phi)^\alpha.  \label{Thm-Phi-Exp}
\end{equation}
{\bf Case 1:} When $0 < \alpha <1$. 

By applying Lemma \ref{Lemma2} to the above expression \eqref{Thm-Phi-Exp} we get, 
\begin{eqnarray}
\sum_{v \in V} p_{f_1}(v)^\alpha &\leq & \sum_{v \in V} (p_{f_2}(v))^\alpha + N\cdot\phi^\alpha.  
\shortintertext{Taking logarithms on either side, we get  }
\log\left(\sum_{v \in V} p_{f_1}(v)^\alpha\right) &\leq&\log\left(\sum_{v \in V} (p_{f_2}(v))^\alpha + N\cdot\phi^\alpha\right), \\ 
&=& \log\left[\sum_{v \in V} (p_{f_2}(v))^\alpha \left(1 + \frac{N\phi^\alpha}{\sum_{v \in V} (p_{f_2}(v))^\alpha} \right)\right],  \\ 
&=& \log\left(\sum_{v \in V} p_{f_2}(v)^\alpha \right) + \log\left(1 + \frac{N\phi^\alpha}{\sum_{v \in V} (p_{f_2}(v))^\alpha} \right), \\ 
&<& \log\left(\sum_{v \in V} (p_{f_2}(v))^\alpha \right) + \left(\frac{N\phi^\alpha}{\sum_{v \in V} (p_{f_2}(v))^\alpha} \right).
\end{eqnarray}
Dividing by $(1- \alpha)$, we get the desired expression \eqref{Eqn7}. 

\noindent{\bf Case 2:} When $\alpha >1$. 

By applying Lemma \ref{Lemma2} to the expression \eqref{Thm-Phi-Exp} we get, 
\begin{eqnarray}
\sum_{v \in V} p_{f_1}(v)^\alpha &\leq& \left[ \left(\sum_{v \in V} (p_{f_2}(v))^\alpha\right)^{1/\alpha} + \left(N\cdot\phi^\alpha\right)^{1/\alpha}\right]^{\alpha}. 
\shortintertext{Now, taking logarithms on either side, we get }
%
\log\left(\displaystyle\sum_{v \in V} p_{f_1}(v)^\alpha\right)
&\leq& \alpha\log\left[ \left(\sum_{v \in V} (p_{f_2}(v))^\alpha\right)^{1/\alpha} + N^{\frac{1}{\alpha}}\cdot\phi\right], \\ 
&=& \alpha\log\left[\left(\sum_{v \in V} (p_{f_2}(v))^\alpha\right)^{\frac{1}{\alpha}} \left(1 + \frac{N^{\frac{1}{\alpha}}\cdot\phi}{\left(\sum_{v \in V} (p_{f_2}(v))^\alpha\right)^{\frac{1}{\alpha}}}\right)\right], \\ 
&=& \log\left(\sum_{v \in V} (p_{f_2}(v))^\alpha \right) + \alpha\log\left(1 + \frac{N^{1/\alpha}\cdot\phi}{\left(\sum_{v \in V} (p_{f_2}(v))^\alpha\right)^{\frac{1}{\alpha}}}\right), \\ 
&<& \log\left(\sum_{v \in V} (p_{f_2}(v))^\alpha \right) + \frac{\alpha \cdot N^{\frac{1}{\alpha}}\cdot\phi}{\left(\sum_{v \in V} (p_{f_2}(v))^\alpha\right)^{\frac{1}{\alpha}}}.
\end{eqnarray}
Dividing by $(1- \alpha)$, we get the desired expression \eqref{Eqn8}. \hfill{$\Box$}

%
%

\begin{theorem}\label{Thm-f_1+f_2} 
Let  $f(v) = c_1f_1(v) +c_2f_2(v)$, $\forall v \in V$.  Then, 
\begin{eqnarray} 
\shortintertext{for $0 <\alpha < 1$, }
H_{\alpha,f}(G)  &<& H_{\alpha,f_1}(G) + \frac{\alpha}{1- \alpha}\log A_1 + \frac{1}{1-\alpha}\frac{A_2^\alpha}{A_1^\alpha} \frac{\displaystyle \sum_{v \in V} (p_{f_2}(v))^\alpha}{\displaystyle \sum_{v \in V} (p_{f_1}(v))^\alpha}, 
\label{Eqn9}
\shortintertext{and for $\alpha > 1$, }
H_{\alpha,f}(G)  &>& H_{\alpha,f_1}(G) - \frac{\alpha}{\alpha-1}\log A_1 - \frac{\alpha}{\alpha-1}\frac{A_2}{A_1}\left(\frac{\displaystyle \sum_{v \in V}(p_{f_2}(v))^\alpha}{\displaystyle \sum_{v \in V}(p_{f_1}(v))^\alpha}\right)^{1/\alpha}.
\label{Eqn10}
\end{eqnarray}
Here, $A_1 = \frac{c_1S_1}{c_1S_1 +c_2S_2}$ and $A_2 = \frac{c_2S_2}{c_1S_1 +c_2S_2}$.
\end{theorem}
\noindent {\bf Proof: } Consider, $f(v) = c_1f_1(v) +c_2f_2(v)$, $\forall v \in V$. 
Now let $S := \displaystyle\sum_{v \in V} f(v) =  c_1\sum_{v \in V}f_1(v) +c_2 \sum_{v \in V}f_2(v) = c_1S_1 +c_2S_2$. 
Next consider, 
\begin{eqnarray}
p_f(v) &=& \frac{f(v)}{S} = \frac{c_1f_1(v) +c_2f_2(v)}{S}, \\ 
			&= & \frac{c_1S_1}{S}p_{f_1}(v)+ \frac{c_2S_2}{S}p_{f_2}(v), \\
			&= & A_1 p_{f_1}(v) +A_2 p_{f_2}(v).
\end{eqnarray} 

Now raising to the power $\alpha$ and summing it over all the vertices of $G$, we get 
\begin{equation}
\sum_{v\in V}p_f(v)^\alpha = \sum_{v \in V} (A_1 p_{f_1}(v) +A_2 p_{f_2}(v))^\alpha 
\end{equation}

Now let us make the cases  of the theorem. 

\noindent {\bf Case 1:  $0 < \alpha <1$.} Applying Lemma \ref{Lemma2}, we get 
\begin{eqnarray}
\sum_{v\in V}p_f(v)^\alpha &=& \sum_{v \in V} (A_1 p_{f_1}(v) +A_2 p_{f_2}(v))^\alpha, \\  
&\leq &  \sum_{v \in V} (A_1 p_{f_1}(v))^\alpha + \sum_{v \in V}(A_2 p_{f_2}(v))^\alpha, \\  
& = &  A_1^\alpha \sum_{v \in V} (p_{f_1}(v))^\alpha + A_2^\alpha\sum_{v \in V}(p_{f_2}(v))^\alpha.
\end{eqnarray}
Taking logarithms on either side, we get 

\begin{eqnarray}
\displaystyle\log\sum_{v\in V}p_f(v)^\alpha 
&\leq & \log \left(A_1^\alpha \sum_{v \in V} (p_{f_1}(v))^\alpha + A_2^\alpha\sum_{v \in V}(p_{f_2}(v))^\alpha\right), \label{Eqn-f_1-f_2_alpha-0-1} \\
& =& \log \left[\left(A_1^\alpha \sum_{v \in V} (p_{f_1}(v))^\alpha\right) \left(1 + \frac{A_2^\alpha\sum_{v \in V}(p_{f_2}(v))^\alpha}{A_1^\alpha \sum_{v \in V} (p_{f_1}(v))^\alpha}\right)\right], \\
& =& \log\left(A_1^\alpha \sum_{v \in V} (p_{f_1}(v))^\alpha\right) +\log \left(1 + \frac{A_2^\alpha\sum_{v \in V}(p_{f_2}(v))^\alpha}{A_1^\alpha \sum_{v \in V} (p_{f_1}(v))^\alpha}\right), \\ 
&< & \alpha \log A_1 + \log\left(\sum_{v \in V} (p_{f_1}(v))^\alpha\right) +\left(\frac{A_2^\alpha\sum_{v \in V}(p_{f_2}(v))^\alpha}{A_1^\alpha \sum_{v \in V} (p_{f_1}(v))^\alpha}\right),
\end{eqnarray}
Dividing by $1-\alpha$, we get the required expression \eqref{Eqn9}.

\noindent {\bf Case 2:  $\alpha >1$.} Applying Lemma \ref{Lemma2} yields,  
\begin{eqnarray}
\sum_{v\in V}p_f(v)^\alpha &=& \sum_{v \in V} (A_1 p_{f_1}(v) +A_2 p_{f_2}(v))^\alpha, \\  
&\leq &  \left[\left(\sum_{v \in V} (A_1 p_{f_1}(v))^\alpha\right)^{\frac{1}{\alpha}} + \left(\sum_{v \in V}(A_2 p_{f_2}(v))^\alpha\right)^{\frac{1}{\alpha}} \right]^\alpha, \\  
& = &  \left[A_1\left(\sum_{v \in V} (p_{f_1}(v))^\alpha\right)^{\frac{1}{\alpha}} + A_2\left(\sum_{v \in V}(p_{f_2}(v))^\alpha\right)^{\frac{1}{\alpha}}\right]^\alpha.
\end{eqnarray}
Taking logarithms on either side, we get 
\begin{eqnarray}
\displaystyle\log \sum_{v\in V}p_f(v)^\alpha &\leq & \alpha\log \left[A_1\left(\sum_{v \in V} (p_{f_1}(v))^\alpha\right)^{\frac{1}{\alpha}} + A_2\left(\sum_{v \in V}(p_{f_2}(v))^\alpha\right)^{\frac{1}{\alpha}}\right], \label{Eqn-f_1-f_2_alpha-1}\\
& =& \alpha \log \left[\left(A_1\left(\sum_{v \in V}(p_{f_1}(v))^\alpha\right)^{\frac{1}{\alpha}}\right) \left(1 + \frac{A_2\left(\sum_{v \in V}(p_{f_2}(v))^\alpha\right)^{\frac{1}{\alpha}}}{A_1\left(\sum_{v \in V}(p_{f_1}(v))^\alpha\right)^{\frac{1}{\alpha}}}\right)\right], \\
& =& \alpha \log\left(A_1\left(\sum_{v \in V}(p_{f_1}(v))^\alpha\right)^{\frac{1}{\alpha}}\right) + \alpha \log \left(1 + \frac{A_2\left(\sum_{v \in V}(p_{f_2}(v))^\alpha\right)^{\frac{1}{\alpha}}}{A_1\left(\sum_{v \in V}(p_{f_1}(v))^\alpha\right)^{\frac{1}{\alpha}}}\right), \\ 
&< & \alpha \log A_1 + \log\left(\sum_{v \in V} (p_{f_1}(v))^\alpha\right) + \frac{\alpha \cdot A_2\left(\sum_{v \in V}(p_{f_2}(v))^\alpha\right)^{\frac{1}{\alpha}}}{A_1\left(\sum_{v \in V}(p_{f_1}(v))^\alpha\right)^{\frac{1}{\alpha}}},
\end{eqnarray}
Dividing by $1-\alpha$, we get the required expression \eqref{Eqn10}. \hfill{$\Box$}
%
%

\begin{cor}
Let  $f(v) = c_1f_1(v) +c_2f_2(v)$, $\forall v \in V$. \\ 
If $0 <\alpha < 1$, then 
\begin{equation}
\begin{split}
H_{\alpha,f}(G) < & \frac{1}{2}[H_{\alpha,f_1}(G) + H_{\alpha,f_2}(G)] +   
\frac{\alpha}{2(1- \alpha)}\log (A_1A_2) + \\ 
& \frac{1}{2(1-\alpha)}\left[\frac{A_2^\alpha}{A_1^\alpha} \frac{\displaystyle \sum_{v \in V}(p_{f_2}(v))^\alpha}{\displaystyle \sum_{v \in V} (p_{f_1}(v))^\alpha} + \frac{A_1^\alpha}{A_2^\alpha} \frac{\displaystyle \sum_{v \in V} (p_{f_1}(v))^\alpha}{\displaystyle \sum_{v \in V} (p_{f_2}(v))^\alpha}\right]. 
\end{split}
\label{Eqn11}
\end{equation}
If  $\alpha > 1$, then  
\begin{equation}
\begin{split}
H_{\alpha,f}(G) > &\frac{1}{2}[H_{\alpha,f_1}(G) + H_{\alpha,f_2}(G)] - \frac{\alpha}{2(\alpha-1)}\log (A_1A_2)  \\ &- \frac{\alpha}{2(\alpha-1)}\left[\frac{A_2}{A_1}\left(\frac{\displaystyle \sum_{v \in V}(p_{f_2}(v))^\alpha}{\displaystyle \sum_{v \in V} (p_{f_1}(v))^\alpha}\right)^{1/\alpha}+\frac{A_1}{A_2}\left(\frac{\displaystyle \sum_{v \in V}(p_{f_1}(v))^\alpha}{\displaystyle \sum_{v \in V} (p_{f_2}(v))^\alpha}\right)^{1/\alpha}\right]. 
\end{split}
\label{Eqn12}
\end{equation}
Here, $A_1 = \frac{c_1S_1}{c_1S_1 +c_2S_2}$ and $A_2 = \frac{c_2S_2}{c_1S_1 +c_2S_2}$. 
\end{cor}
\noindent \textbf{Proof:} The proof follows similar to Theorem \ref{Thm-f_1+f_2}. In the case of  $0 \leq \alpha <1$, the equation \eqref{Eqn-f_1-f_2_alpha-0-1} is expressed as follows:
\begin{align}
\displaystyle\log\sum_{v\in V}p_f(v)^\alpha \leq & \log \left(A_1^\alpha \sum_{v \in V} (p_{f_1}(v))^\alpha + A_2^\alpha\sum_{v \in V}(p_{f_2}(v))^\alpha\right), \\ 
\begin{split}
 = &  \frac{1}{2}\log \left(A_1^\alpha \sum_{v \in V} (p_{f_1}(v))^\alpha + A_2^\alpha\sum_{v \in V}(p_{f_2}(v))^\alpha\right) \\& \quad  + \frac{1}{2}\log \left(A_1^\alpha \sum_{v \in V} (p_{f_1}(v))^\alpha + A_2^\alpha\sum_{v \in V}(p_{f_2}(v))^\alpha\right), \end{split} \\
\begin{split}
 = & \frac{1}{2}\log \left[\left(A_1^\alpha \sum_{v \in V} (p_{f_1}(v))^\alpha\right) \left(1 + \frac{A_2^\alpha\sum_{v \in V}(p_{f_2}(v))^\alpha}{A_1^\alpha \sum_{v \in V} (p_{f_1}(v))^\alpha}\right)\right] \\
 & \quad + \frac{1}{2}\log \left[\left(A_2^\alpha \sum_{v \in V} (p_{f_2}(v))^\alpha\right) \left(1 + \frac{A_1^\alpha\sum_{v \in V}(p_{f_1}(v))^\alpha}{A_2^\alpha \sum_{v \in V} (p_{f_2}(v))^\alpha}\right)\right]\end{split}
\end{align}

Now proceeding as before and by simplifying each of the terms in the above equation, we get the required expression \eqref{Eqn11}.

Similarly, in the case of $\alpha >1$, the expression \eqref{Eqn-f_1-f_2_alpha-1} is expressed as,  
\begin{align}
\displaystyle\log \sum_{v\in V}p_f(v)^\alpha  \leq & \alpha\log \left[A_1\left(\sum_{v \in V} (p_{f_1}(v))^\alpha\right)^{\frac{1}{\alpha}} + A_2\left(\sum_{v \in V}(p_{f_2}(v))^\alpha\right)^{\frac{1}{\alpha}}\right],\\
\begin{split}
= &  \frac{\alpha}{2}\log \left[A_1\left(\sum_{v \in V} (p_{f_1}(v))^\alpha\right)^{\frac{1}{\alpha}} + A_2\left(\sum_{v \in V}(p_{f_2}(v))^\alpha\right)^{\frac{1}{\alpha}}\right]\\
 & \quad +  \frac{\alpha}{2}\log \left[A_1\left(\sum_{v \in V} (p_{f_1}(v))^\alpha\right)^{\frac{1}{\alpha}} + A_2\left(\sum_{v \in V}(p_{f_2}(v))^\alpha\right)^{\frac{1}{\alpha}}\right], 
\end{split} \\
\begin{split}
 = & \frac{\alpha}{2} \log \left[\left(A_1\left(\sum_{v \in V}(p_{f_1}(v))^\alpha\right)^{\frac{1}{\alpha}}\right) \left(1 + \frac{A_2\left(\sum_{v \in V}(p_{f_2}(v))^\alpha\right)^{\frac{1}{\alpha}}}{A_1\left(\sum_{v \in V}(p_{f_1}(v))^\alpha\right)^{\frac{1}{\alpha}}}\right)\right] \\ 
 & \quad +  \frac{\alpha}{2} \log \left[\left(A_2\left(\sum_{v \in V}(p_{f_2}(v))^\alpha\right)^{\frac{1}{\alpha}}\right) \left(1 + \frac{A_1\left(\sum_{v \in V}(p_{f_1}(v))^\alpha\right)^{\frac{1}{\alpha}}}{A_2\left(\sum_{v \in V}(p_{f_2}(v))^\alpha\right)^{\frac{1}{\alpha}}}\right)\right] \end{split} 
\end{align}

As before upon simplification of the above equation, we get the required expression \eqref{Eqn12}. \hfill{$\Box$}

\subsection*{Applications to chemical graphs}

In this section, we consider various classes of simple chemical graphs and illustrate the results from previous section. To this purpose we consider a specific example of the equivalence relation $\Gamma$ on $G$ and an information functional $f_P$. In order to define concrete graph entropies, we need to specify graph invariants and information functionals to determine a probability distribution. 

For the graph invariant we use the automorphism group of a graph. We use this invariant due to their extensive investigations available in the literature; for example see \cite{chLS:Mowshowitz1}. Note that there are various other invariants such as distance, degrees and paths that could be used.  Observe that each graph belongs to an automorphism group, where an automorphism is a permutation of the vertices such that the adjacency relation of the graph is preserved. An automorphism group divides the vertex set into orbits where  a vertex orbit is a collection of topologically equivalent vertices. 

\begin{defn}
Let $\Gamma$ be an automorphism (equivalence relation) that partitions the vertex set $V$ of $G$ into vertex orbits.  Let $X_1, \ldots, X_k$ be the $k$ orbits of $V$ such that $|V| = |X_1| + \cdots + |X_k|$. 
\end{defn}

For the information functional, we reproduce the definitions of two information functionals based on metrical properties of graphs \cite{chLS:Dehmer1,dehmer_mowshowitz_2010,skorobogatov}.

Let $G=(V,E)$ be a simple, undirected graph on $n$ vertices and let $d(u,v)$ denote the distance between two vertices $u$ and $v$, and let $\eta(G) = \max\{d(u,v): u,v\in V\}$. Let $S_j(u;G)$ denote the $j$-sphere of a vertex $u$ defined as $S_j(u;G) = \{x \in V: d(u,x) = j\}$. 

\begin{defn}
Parameterized linear information functional using $j$-spheres:
\begin{equation}\label{fn.al-lin-j-sphere}
f_P'(v_i) = \sum_{j=1}^{\eta(G)} c_j |S_j(v_i; G)|,
\end{equation}
where $c_k > 0$ for $1\leq k \leq \eta(G)$.
\end{defn}

\begin{defn}
Parameterized exponential information functional using $j$-spheres:
\begin{equation}\label{fn.al-exp-j-sphere}
f_P(v_i) = \displaystyle \beta^{\sum_{j=1}^{\eta(G)} c_j |S_j(v_i;G)| }, 
\end{equation}
where $\beta > 0$  and $c_k > 0$ for $1\leq k \leq \eta(G)$.
\end{defn}
Note that the constants $c_j$'s in the above expressions contribute to the weight of the $j$-spheres and hence the functionals could be meaningfully interpreted when the $c_j$'s are all distinct.

\subsection*{Stars}

A Star $S_n$ is a tree on $n$ vertices where there is exactly one vertex of degree $n-1$ and $n-1$ vertices of degree $n$.  
Star graphs have been of considerable interest, since they represent trees with smallest possible diameter among all trees on $n$ vertices.  
\begin{theorem} \label{Thm-Star-Gamma}
Let $\Gamma$ be an automorphism defined on $S_n$ such that $\Gamma$ partitions $V(S_n)$ into two orbits, $X_1$ and $X_2$, with $|X_1| = 1$ and $|X_2| = n-1$.   Then 
\begin{eqnarray}
\shortintertext{for $ 0 <\alpha <1$, } 
H_{\alpha,\Gamma}(S_n) &<& \log n - \frac{n-1}{n} \log(n-1) + \frac{(1-\alpha)(n-1)^{\alpha-2}}{\ln 2},
\shortintertext{and for $\alpha >1$, }
H_{\alpha,\Gamma}(S_n) &>& \log n - \frac{n-1}{n} \log(n-1) - \frac{\alpha-1}{(n-1)^{\alpha-2}\ln 2}.
\end{eqnarray}
\end{theorem}
\noindent \textbf{Proof:} Let $p_1 = \frac{|X_1|}{|V|} = \frac{1}{n}$ and $p_2 = \frac{|X_2|}{|V|} = \frac{n-1}{n}$. So, $\rho = \max\{\frac{p_1}{p_2},\frac{p_2}{p_1} \} = n-1$.  Now, we have 
\begin{eqnarray}
H_{\Gamma}(S_n) &=& \log_2 n - \frac{n-1}{n} \log_2 (n-1). 
\shortintertext{Observe that, }
H_{\alpha, \Gamma}(S_n) &=& \frac{1}{1-\alpha}\left[ \log_2 (1+(n-1)^\alpha ) - \alpha \log_2 n\right]. \label{Eqn-Sn-alpha-gamma}
\end{eqnarray}
Now by Theorem \ref{Thm-alpha-Gamma-Gamma}, we have 
\begin{eqnarray}
H_{\alpha, \Gamma}(S_n) &<& H_{\Gamma}(S_n) + \frac{2 (2-1)(1-\alpha) \rho^{\alpha-2}}{2 \cdot \ln 2},
\shortintertext{for  $0 \leq \alpha \leq 1$. Hence,}
H_{\alpha, \Gamma}(S_n)&<&  \log_2 n - \frac{n-1}{n} \log_2 (n-1)+ \frac{(1-\alpha) (n-1)^{\alpha-2}}{\ln 2}.
\end{eqnarray}
Similarly, for $\alpha >1$, we have  by Theorem \ref{Thm-alpha-Gamma-Gamma}, 
\begin{eqnarray}
H_{\alpha, \Gamma}(S_n) &>& H_{\Gamma}(S_n) - \frac{2 (2-1)(\alpha-1)}{2 \cdot \ln 2\cdot \rho^{\alpha-2}}.
\shortintertext{That is, }
H_{\alpha, \Gamma}(S_n) &>&  \log_2 n - \frac{n-1}{n} \log_2 (n-1)- \frac{(\alpha-1)}{(n-1)^{\alpha-2} \cdot \ln 2}.
\end{eqnarray}
Hence the theorem. \hfill{$\Box$}

\begin{theorem}\label{Thm-Star-f}
 Let $\Gamma$ be an automorphism on $V(S_n)$ and let $f$  be any information functional defined on $V(S_n)$ such that $|X_1| < f(v_i)$ and $|X_2| < f(v_j)$ for some $i$ and  $j$, $1\leq i \neq j \leq n$. Then, for $ 0 <\alpha <1$, 
\begin{eqnarray}
H_{\alpha,f}(S_n) & > & \frac{1}{1- \alpha}\log (1+(n-1)^\alpha) - \frac{\alpha}{1-\alpha} \log S, \label{Eqn-Star-f-1}
\shortintertext{and for $\alpha >1$, } 
H_{\alpha,f}(S_n) & < &\frac{1}{1-\alpha}\log (1+(n-1)^\alpha) + \frac{\alpha}{\alpha-1} \log S. \label{Eqn-Star-f-2}
\end{eqnarray}
Here $S = \sum_{v \in V} f(v)$.
\end{theorem}
\noindent \textbf{Proof:} Follows by using equation \eqref{Eqn-Sn-alpha-gamma} in Theorem \ref{Thm-Gamma-f}. \hfill{$\Box$}
\begin{remark}{\rm
For instance, when $f = f_P'$ defined by equation \eqref{fn.al-lin-j-sphere} with constants $c_1, c_2 \geq 1$. We have $S = \sum_{v \in V(S_n)} f(v) =   (2c_1 +c_2(n-2))(n-1)$. By substituting the value of $S$ in equations \eqref{Eqn-Star-f-1} and \eqref{Eqn-Star-f-2}, we get the bounds for $H_{\alpha, f_P'}(S_n)$.   }
\end{remark}


\begin{remark} {\rm
At this point, observe that another class of graph that possesses the same automorphism group as the stars is the wheel graph. A wheel $W_n$ is a graph obtained by joining a new vertex $v$ to every vertex of an $(n-1)$-cycle $C_{n-1}$. That is, $W_n = C_{n-1} +\{v\}$. While studying the inequalities for this class of graph, we derived similar expressions as of theorems \ref{Thm-Star-Gamma} and \ref{Thm-Star-f}. Hence we conclude that the theorems \ref{Thm-Star-Gamma} and \ref{Thm-Star-f} also holds for the wheel $W_n$. }
\end{remark}

\subsection*{Paths}
A path graph, denoted by $P_n$, are the only trees with maximum diameter among all the trees on $n$ vertices. This class of graph has received considerable attention in chemistry when studying the hydrogen-depleted hydrocarbon molecules. Let $\Gamma$ be an automorphism defined on $P_n$, where $\Gamma$ partitions the vertices of $P_n$ into   $\frac{n}{2}$ orbits ($X_i$) of size 2, when $n$ is even, and  $\frac{n-1}{2}$ orbits of size 2 and one orbit of size 1, when $n$ is odd.

\begin{theorem}
Let $f$ be any information functional such that $f(v) > 2 $ for at least $\frac{n}{2}$ vertices of $P_n$ and let $\Gamma$ be as defined above. Then   
\begin{eqnarray}
H_{\alpha, \Gamma}(P_n) & = &\log_2 \frac{n}{2},  \\
H_{\alpha, f}(P_n) &>& \frac{1}{1-\alpha}\log_2 n - \frac{\alpha}{1-\alpha}\log_2 S  - 1,  \text{ if } 0 < \alpha < 1, \\
H_{\alpha, f}(P_n) &< & \frac{1}{1-\alpha}\log_2 n + \frac{\alpha}{\alpha-1}\log_2 S  - 1, \text{ if } \alpha > 1,
\end{eqnarray}
where $S = \sum_{v \in V} f(v)$.
\end{theorem}

\noindent{\bf Proof:} $H_{\alpha, \Gamma}(P_n)$ follows directly from the definition of R\'{e}nyi entropy. 
Next, by using the value of $H_{\alpha, \Gamma}(P_n)$ in Theorem \ref{Thm-Gamma-f}, we get the desired expression for $H_{\alpha,f}(P_n)$. \hfill{$\Box$}

\subsection*{Connected graphs}
In this section, we consider any general connected graph $G$ and the functionals $f_P$ and $f_P'$ given by equations \eqref{fn.al-exp-j-sphere} and \eqref{fn.al-lin-j-sphere} respectively.
In the next two theorems, we present the explicit bounds for the R\'{e}nyi entropy $H_{\alpha,f}(G)$, when we choose the two information functionals in particular. 
 
\begin{theorem} Let $f = f_P'$ given by equation \eqref{fn.al-lin-j-sphere}. Let $c_{\max} = \max\{c_i: 1\leq i \leq \eta(G)\}$ and $c_{\min} = \min\{c_i: 1\leq i \leq \eta(G)\}$ where $c_i$ is defined in $f_P'$. Then the value of $H_{\alpha, f_P'}(G)$ lies within the following bounds. 
\begin{eqnarray}
\shortintertext{When $0 < \alpha <1,$ }
\log_2 n - \frac{\alpha}{1- \alpha} \log_2 \frac{c_{\max}}{ c_{\min} } & \leq H_{\alpha, f_P'}(G) \leq & \log_2 n + \frac{\alpha}{1- \alpha} \log_2 \frac{c_{\max}}{c_{\min}}, \label{Eqn13}   \shortintertext{and when $\alpha >1$,}
\log_2 n - \frac{\alpha}{\alpha-1} \log_2 \frac{c_{\max}}{c_{\min}} &\leq H_{\alpha, f_P'}(G) \leq &\log_2 n + \frac{\alpha}{\alpha-1} \log_2 \frac{c_{\max}}{c_{\min}}. \label{Eqn14} 
\end{eqnarray}
\end{theorem}

\noindent {\bf Proof:} Given $f(v) = f_P'(v) = \sum_{j=1}^{\eta(G)} c_j|S_j(v;G)|$ with $c_j > 0$ for $1\leq j \leq \eta(G)$. Let $c_{\max} = \max \{c_j : 1\leq j \leq \eta(G)\}$ and $c_{\min} = \min \{c_j : 1\leq j \leq \eta(G)\}$. We have,
\begin{eqnarray}
f(v) &= &\displaystyle\sum_{j=1}^{\eta(G)} c_j |S_j(v;G)| \leq   (n-1)c_{\max}. \label{upperbound}
\shortintertext{ Similarly, }
f(v) & \geq & (n-1)c_{\min}. \label{lowerbound}
\end{eqnarray}
Therefore, combining the Equations~\eqref{upperbound} and \eqref{lowerbound} and by adding over all the vertices of $G$, we get
\begin{eqnarray}
n (n-1) c_{\min} \leq & \displaystyle \sum_{v\in V} f(v) &\leq  n (n-1) c_{\max}.
\shortintertext{ Hence,}
\displaystyle \frac{c_{\min}}{n\cdot c_{\max}}  \leq & p_f(v)& \leq  \frac{c_{\max}}{n\cdot c_{\min}}. 
\shortintertext{By raising to the power $\alpha$ and adding over all the vertices of $G$, we have} 
\displaystyle  n \cdot \left(\frac{c_{\min}}{n\cdot c_{\max}}\right)^\alpha  \leq & \displaystyle \sum_{v \in V}p_f(v)^\alpha & \leq  n \cdot \left(\frac{c_{\max}}{n\cdot c_{\min}}\right)^\alpha. 
\shortintertext{Taking logarithms we get,}
\displaystyle \log_2 n + \alpha \log_2\frac{c_{\min}}{n\cdot c_{\max}}  \leq & \displaystyle\log_2\left(\sum_{v \in V}p_f(v)^\alpha \right) & \leq  \log_2 n + \alpha \log_2 \frac{c_{\max}}{n\cdot c_{\min}}. \label{Eqn-bounds-Renyi-Prob}
\end{eqnarray}

Dividing the expression \eqref{Eqn-bounds-Renyi-Prob} by $(1-\alpha)$, and simplifying we get the desired expressions given by \eqref{Eqn13} and \eqref{Eqn14} depending on the value of $\alpha$. ~\hfill{$\Box$}
 
\begin{theorem}
Let $f = f_P$ given by equation \eqref{fn.al-exp-j-sphere}. Let $c_{\max} = \max\{c_i: 1\leq i \leq \eta(G)\}$ and $c_{\min} = \min\{c_i: 1\leq i \leq \eta(G)\}$ where $c_i$ is as defined in $f_P$. Then the value of $H_{\alpha, f_P'}(G)$ can be bounded as follows.
\begin{eqnarray}
\shortintertext{When $0 < \alpha <1$, }
\log_2 n - \frac{\alpha (n-1)X}{1- \alpha} \log_2 \beta & \leq H_{\alpha, f_P}(P_n) \leq & \log_2 n + \frac{\alpha (n-1)X}{1- \alpha} \log_2 \beta, \label{Eqn15}
\shortintertext{and when $\alpha >1$,}
\log_2 n -\frac{\alpha (n-1)X}{\alpha - 1} \log_2 \beta & \leq H_{\alpha, f_P}(P_n) \leq & \log_2 n + \frac{\alpha (n-1)X}{\alpha -1} \log_2 \beta, \label{Eqn16}
\end{eqnarray}
where $X = c_{\max}- c_{\min}$. 
\end{theorem}

\noindent {\bf Proof:} Given $f(v) = f_P(v) = \displaystyle \beta^{\sum_{j=1}^{\eta(G)} c_j|S_j(v;G)|}$ with $c_j > 0$ for $1\leq j \leq \eta(G)$. Let $c_{\max} = \max \{c_j : 1\leq j \leq \eta(G)\}$ and $c_{\min} = \min \{c_j : 1\leq j \leq \eta(G)\}$. We have,
\begin{eqnarray}
f(v) &= &\displaystyle \beta^{\sum_{j=1}^{\eta(G)} c_j |S_j(v;G)|} \leq  \beta^{(n-1)c_{\max}}. \label{upperbound-f_P}
\shortintertext{ Similarly, }
f(v) & \geq & \beta^{(n-1)c_{\min}}. \label{lowerbound-f_P}
\end{eqnarray}
Therefore, combining the Equations~\eqref{upperbound-f_P} and \eqref{lowerbound-f_P} and adding over all the vertices of $G$, we get
\begin{eqnarray}
n \cdot\beta^{(n-1) c_{\min}} \leq & \displaystyle \sum_{v\in V} f(v) &\leq  n \cdot \beta^{(n-1) c_{\max}}.
\shortintertext{ Hence,}
\displaystyle \frac{\beta^{(n-1)(c_{\min} - c_{\max})}}{n}  \leq & p_f(v)& \leq  \frac{\beta^{(n-1)(c_{\max} - c_{\min})}}{n}. 
\end{eqnarray}
Let $X= c_{\max} - c_{\min}$. Now, by raising $p_f(v)$ to the power $\alpha$ and adding over all the vertices of $G$, we have,  
\begin{eqnarray}
\displaystyle  n \cdot \left(\frac{1}{n \cdot \beta^{(n-1)X}}\right)^\alpha  \leq & \displaystyle \sum_{v \in V}p_f(v)^\alpha & \leq  n \cdot \left(\frac{\beta^{(n-1)X}}{n}\right)^\alpha. 
\shortintertext{Taking logarithms we get,}
\displaystyle \log_2 n - \alpha \log_2 (n \cdot \beta^{(n-1)X})  \leq & \displaystyle\log_2\left(\sum_{v \in V}p_f(v)^\alpha \right) & \leq  \log_2 n + \alpha \log_2\left(\frac{\beta^{(n-1)X}}{n}\right). \\
\displaystyle (1- \alpha)\log_2 n - \alpha (n-1) X \log_2 \beta  \leq & \displaystyle\log_2\left(\sum_{v \in V}p_f(v)^\alpha \right) & \leq  (1- \alpha)\log_2 n + \alpha (n-1) X \log_2 \beta. 
 \label{Eqn-bounds-Renyi-Prob-f_P}
\end{eqnarray}

Dividing the expression \eqref{Eqn-bounds-Renyi-Prob-f_P} by $(1-\alpha)$, and simplifying we get the desired expressions given by \eqref{Eqn15} and \eqref{Eqn16}. ~\hfill{$\Box$}





\subsection*{Conclusions and Summary}

In this article, we have studied one of the most daunting problems in the study of information measures, that is, establishing relations between graph entropy measures. Among all the entropy measures, we have considered the graph entropies defined using the classical Shannon entropy and the R\'{e}nyi entropy. Further, we have also considered two major types of probability distributions in the definitions of the entropies, namely the classical partition-based distributions and the recent non-partition-based distribution defined on the vertices of the graph by using information functionals. Thus, we have established analytical relations when the two different types of distribution are used for the measures. 

In general, the results could be used in various branches of science
including mathematics, statistics, information theory, biology, chemistry and social sciences. Further, the determination of analytical relations between measures gains 
practical importance when dealing with large scale networks. Moreover, relations involving quantitative network measures could be fruitful when determining the information content of large complex networks.


\section*{Acknowledgments}

Matthias Dehmer and Lavanya Sivakumar thank the FWF (Project No. PN22029-N13) for supporting this work.

\bibliography{Lanybib-InfoTheory,matthias_bibtex_current}



\end{document}